\documentclass[aps,pra,pdf,superscriptaddress,amsmath,amssymb,amsfonts,twocolumn,showpacs,nofootinbib]{revtex4-1}

\usepackage{amssymb}
\usepackage{eqnarray,amsmath}
\newcommand{\abs}[1]{\left| #1 \right|} 
\usepackage{graphicx}
\usepackage{epsfig}
\usepackage{dcolumn}
\usepackage{bm}
\usepackage{braket}
\usepackage{amsmath}
\usepackage{hyperref}
\usepackage{physics}
\usepackage{mathtools}
\usepackage{graphicx,color,xcolor,colortbl}
\hypersetup{colorlinks=true, citecolor=blue, urlcolor=blue, linkcolor=blue}
\usepackage[normalem]{ulem} %% REMOVE ME!

\definecolor{Gray}{gray}{0.85}
\definecolor{LightCyan}{rgb}{0.88,1,1}

\newcolumntype{a}{>{\columncolor{Gray}}c}
\newcolumntype{b}{>{\columncolor{white}}c}

\begin{document}

\title{Supersolid Stacks in Antidipolar Bose-Einstein Condensates}

\author{K. Mukherjee}
\author{M. Nilsson Tengstrand}
\author{T. Arnone Cardinale}
\author{S. M. Reimann}
\affiliation{Mathematical Physics and NanoLund, LTH, Lund University, Box 118, 22100 Lund, Sweden}

\begin{abstract}
We theoretically investigate a novel supersolid structure  taking the form of stacked, disk-shaped superfluid droplets connected via a dilute superfluid, in an antidipolar condensate. A phase diagram is  determined for varying the particle number and scattering length, identifying the regions of a regular dipolar superfluid, supersolid stacks, and isolated stacked disk-shaped droplets in an experimentally realizable trapping potential. The collective   Bogoliubov excitation spectrum across the superfluid-supersolid phase transition is studied, and the transition point is found to be associated with the breaking of the degeneracy of the two lowest-lying modes. The dynamical generation of the supersolid stacks is also investigated by ramping down the scattering length across the phase transition. Moreover, we have studied the impact of vortex-line penetration on the phase transition.  We have found that the presence of a vortex line causes the supersolid region to move towards weaker contact interactions. Our detailed numerical simulations highlight that an antidipolar condensate can create such supersolid stacks within an experimentally reachable parameter regime.
\end{abstract}
%\pacs{47.20.Dr, 47.35.Pq, 47.54.-r}
\date{\today}

\maketitle
\section{Introduction}
Dipolar Bose-Einstein condensates offer an intriguing platform~\cite{Baranov2008,Lahaye2009,Baranov2012,Boettcher2021,Chomaz2022} to study 
the supersolid state of matter~\cite{Boettcher2019,Tanzi2019a,Chomaz2019,Natale2019,Tanzi2019b,Guo2019,Hertkorn2019,Hertkorn2021b}, displaying both diagonal and off-diagonal long-range order~\cite{Gross1957,*Gross1958,Yang1962,Andreev1969,Chester1970,Leggett1970,Pomeau1994,Boninsegni2012}.
In a supersolid, translational symmetry is spontaneously broken, leading to density modulations 
while   partially maintaining superfluid properties.  
The concept was originally introduced and debated over an extensive  period of time in the context of $^4$He~\cite{Kim2004a,*Kim2004b,Balibar2010,Kim2012,Boninsegni2012,Chan2013},  however,  only recently ultra-cold atoms have emerged as a viable alternative setup~\cite{Henkel2010,Cinti2010,Saccani2011,Leonard2017a,*Leonard2017b,Lin2011,Li2016,Li2017}. Experiments involving highly magnetic dysprosium~\cite{Kadau2016,Ferrier2016,Schmitt2016,Boettcher2019b} and erbium atoms~\cite{Chomaz2016,Chomaz2018}  uncovered the emergence of supersolidity in (quasi-)one~\cite{Tanzi2019a,Boettcher2019b, Chomaz2018, Sohmen2021} as well as two spatial dimensions~\cite{Schmidt2021,Biagioni2022,Bland2022}. The underlying mechanism is attributed to the interplay between inter-particle interactions and quantum fluctuations~\cite{Lima2011,Wachtler2016a,*Wachtler2016b,Bisset2016}, a mechanism similar to the one seen in binary Bose gases forming droplets~\cite{Petrov2015, Petrov2016, Arlt2018} which also have been realized experimentally~\cite{Cabrera2017, Semeghini2018, Skov2021}. Owing to the unique and fascinating properties of dipolar droplets and supersolids, there has been a recent surge in research focused on understanding these systems \cite{Bisset2015,Xi2016,Blakie2016,Saito2016,Bisset2016,Macia2016,Baillie2017,Edler2017,Baillie2018,Roccuzzo2019,Zhang2019,
	Blakie2020, Mishra2020, Chomaz2020,Poli2021,Hertkorn2021,Hertkorn2021b, Zhang2021, Young2022, Ghosh2022,Gallemi2022,Schmidt2022,Halder2022,Tengstrand2021, Roccuzzo2022}, including the exploration of out-of-equilibrium dynamics~\cite{Sohmen2021,Ilzhofer2021,Norcia2022, Mukherjee2022}, vortices~\cite{Roccuzzo2020, Gallemi2020, Sindik2022}, and extensions to dipolar mixtures~\cite{Smith2021,Bisset2021,Li2022, Scheiermann2023,Bland2022b, Halder2022b}.

The bulk of current and recent work set focus on dipole-dipole interactions under a fixed magnetic field, oriented along a particular direction, say, the $z$-axis. However, by rotating the polarizing magnetic field (a technique also demonstrated in experiments~\cite{Tang2021}) it is possible to manipulate the dipole-dipole interaction~\cite{Giovanazzi2002, Baillie2020}. A time-averaged DDI can be utilized when the rotation frequency exceeds the trap frequencies but is still smaller than the Larmor frequency. The magnitude and polarity of this DDI is determined by the angle between the dipole and the $z$-axis~\cite{Prasad2019}. By adjusting the tilt angle, the effective interaction can be modulated from dipolar to antidipolar regimes where the dipolar interaction is reversed. In this case, head-to-tail antidipoles repel, while side-by-side antidipoles attract, which is just opposite to the behavior observed in the ordinary dipolar regime.
In a single trapped antidipolar condensate, stacks of disk-like droplets may form that can even be supersolid when the disks connect by a dilute superfluid. These structures are  quite different 
from the usual linear arrays of elongated filaments found in conventional dipolar condensates, 
as mentioned early-on in the PhD work of Wenzel~\cite{WenzelPhD}. Stacked droplets or supersolids however have, to the best of our knowledge, not yet gained much attention, despite the in principle realistic experimental scenario, and the many interesting future prospects concerning for example studies of vorticity in such systems. 
Similar stacked supersolid structures were  previously  only discussed 
for a different setting of dipolar and non-dipolar mixtures~\cite{Kirkby2023}, where the immiscibility of the components 
stabilized the system. Other  theoretical~\cite{Ghosh2022, Bilitewski2023} or  experimental studies~\cite{Natale2022, Du2023} with focus on  layered structures of dipolar gases required an optical lattice potential~\cite{Natale2022, Bilitewski2023, Du2023} or an electric dipole moment~\cite{Ghosh2022} for their fabrication. 

Here, we set focus on the antidipolar single-component condensates
where remarkably, the rather unique stack structures can arise exclusively from the combined effects of long-range and contact forces~\cite{WenzelPhD}. The layers can be completely isolated from each other, or be connected by a superfluid link, forming a novel supersolid that we in the following  refer to as a ``supersolid stack". 
We determine the ground-state phase diagram depicting the superfluid (SF) phase, supersolid stack (SSS), and isolated stacked droplets (SDL) as a function of particle number and scattering length, with a fixed antidipolar length. After demarcating the explicit phase boundaries, we proceed to analyze the low-lying collective excitation spectra across the transition using a linear stability analysis based on the Bogoliubov de-Gennes (BdG) approach~\cite{Wilson2009a, Wilson2009b, Lu2010, Blakie2012, Martin2012, Hertkorn2019}. A crucial finding is that the degeneracy of the two lowest modes is broken at the transition point, serving as an indicator of the SF-SSS phase transition. Additionally, we provide evidence that the generation of layered structures can be achieved dynamically from a superfluid state by performing an interaction quench. Such a quenching technique for supersolid state generation is frequently used for regular dipolar condensates~\cite{Tanzi2019a, Boettcher2019, Bland2022}. Quenching the interaction across the phase boundary results in density oscillations that we have found to be associated with patterns in the low-lying excitation spectra of the system. Finally, we investigate the impact of imposing a vortex line through the stacks. Our analysis reveals that a higher-charge vortex shifts the supersolid phase towards lower scattering lengths relative to the system without a vortex. Intriguingly, this suggests that the presence of a vortex affects the superfluid connection and reduces the density of the crystal structure.
  
The remainder of the paper is structured as follows. Section~\ref{model} describes the extended Gross-Pitaevskii (eGP) equation and the BdG approach. In Sec.~\ref{GroundState_sec}, we present the ground state properties of an antidipolar condensate, delineating the phase diagram in Sec.~\ref{phases_subsec}, and describing the collective excitation properties in Sec.~\ref{excitation_subsec}. The dynamical formation of a layered supersolid structure is addressed in Sec.~\ref{dyn_density_profile}, and collective oscillations in Sec.~\ref{collect_osc_dyn}. The  effect of a vortex line on supersolidity is discussed in Sec.~\ref{vortex_sec}. We provide a summary of our findings, along with future perspectives, in Sec.~\ref{conclusions}. In Appendix~\ref{threeBody_appen}, we discuss the impact of a three-body loss term. Appendix~\ref{comdetails} provides  some  details  of our numerical simulations.

\section{MODEL AND METHODS}\label{model}
 \begin{figure*}
 	\centering
 	\includegraphics[width = 0.95\textwidth]{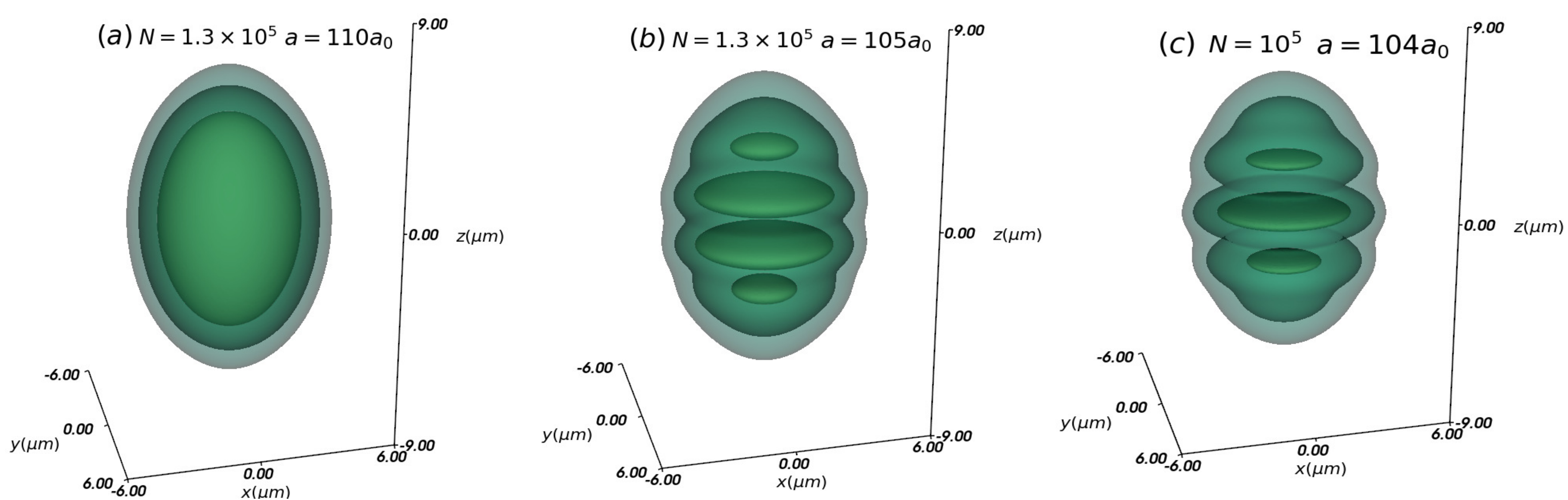}
 	\caption{({\it Color online}) Density isosurfaces representing  (a) the regular superfluid and (b)-(c) the supersolid state formed by layers of four (b) and three (c) disks for a few selected particle numbers $N$ and scattering lengths $a$ (see the legends). The density isosurfaces are taken at $50\%$, $16\%$, and $2.5\%$ of the maximum density. The system is confined in a harmonic trapping potential with frequencies $(\omega_x, \omega_y, \omega_z)/(2\pi) = (100, 100, 50)$, and dipolar length $a_{\rm dd} = -65.5a_{0}$. }\label{den_profile}
 \end{figure*}
\subsection{Extended Gross-Pitaevskii Equation}
We consider a dipolar BEC (dBEC) composed of atoms with mass $M$ and magnetic dipole moment $\mu_m$. The dBEC is confined within a three-dimensional harmonic trapping potential of the form $V(\mathbf{r}) = M(\omega_x^2 x^2 + \omega_y^2 y^2 + \omega_z^2 z^2)/2$. The atomic dipoles are aligned with a rotating uniform magnetic field of strength $\mathbf{B}(t)$ along $\mathbf{e}(t)$, where $\mathbf{e}(t) = \mathbf{B}(t)/|\mathbf{B}(t)|$~\cite{Prasad2019, Baillie2020}. If the external magnetic field rapidly rotates at an angle $\phi$ relative to the $z$-axis and with a much higher frequency than the trap frequency, it is possible to perform a time-averaging process on the dipole-dipole interaction (DDI). The time-averaged DDI is given by~\cite{Giovanazzi2002}
\begin{eqnarray}\label{DDI}   
 U_{\mathrm{dd}} (\textbf{r},t) = \frac{\mu_0 \mu^2_{m}}{4\pi}\left( \frac{3 \cos^2\phi - 1}{2} \right) \left(\frac{1-3\cos^2\theta}{|\vb{r}|^3}\right),
\end{eqnarray}
where $\theta$ denotes the angle between  $\vb{r}$ and the $z$-axis.
The  DDI is highly dependent upon the orientation of the constituent dipoles. The interaction vanishes entirely at the so-called magic angle, denoted by $\phi_{m}\approx 54.7^{\circ}$. Specifically, for $\phi < \phi_{m}$, the preferred orientation of the dipoles is head-to-tail, while for $\phi > \phi_{m}$, an antidipolar configuration with a side-by-side arrangement of dipoles becomes energetically favorable.
At  zero  temperature, the  system  is well described by the eGP equation 
~\cite{Lima2011,Wachtler2016a,*Wachtler2016b, Chomaz2016}
 \begin{eqnarray}\label{eGPE}   
 & i\hbar \partial_t \psi(\textbf{r},t)  =  \bigg[-\frac{\hbar^2}{2M}\nabla^2 + V(\textbf{r}) + g \abs{\psi(\textbf{r},t)}^2+ \frac{3}{4 \pi} g_{\rm dd}\times \nonumber\\&  \int d\textbf{r}^{\prime} \frac{1 -3\cos^{2}\theta}{\abs{\textbf{r} - \textbf{r}'}^3}\abs{\psi(\textbf{r}^{\prime},t)}^2 + \gamma(\epsilon_{\mathrm{dd}})\abs{\psi(\textbf{r},t)}^3 \bigg] \psi(\vb{r},t). 
 \end{eqnarray}
 Here, the short-range repulsive contact interaction,  $g = 4 \pi \hbar^2a/M$, is determined by the $s$-wave scattering length $a$. The dipolar interaction coefficient is $g_{\rm dd} = 4\pi\hbar^2a_{\rm dd}/M$ with  $a_{\rm dd}= \mu_{0}\mu^2_{m}M(3 \cos^2\phi -1)/24\pi \hbar^2$ being the so-called dipolar length. The final term in Eq.~\eqref{eGPE} is given by the repulsive Lee-Huang-Yang (LHY) correction 
 with $\gamma(\epsilon_{\rm dd}) = \frac{32}{3}g \sqrt{\frac{a^3}{\pi}} \left(1+\frac{3}{2}\epsilon_{\rm dd}^2\right)$~\cite{Lima2011, Lima2012}, where the dimensionless parameter $\epsilon_{\rm dd} = a_{\mathrm{dd}}/a$ quantifies the relative strength of the DDI compared to the contact interaction. The solution of Eq.~\eqref{eGPE} is obtained by employing the split-step Crank-Nicholson method~\cite{crank_nicolson1947,Antoine2013} in imaginary time to determine the initial ground state, and in real time to monitor the system's dynamics. The behavior of the system can be classified into different phases depending on the absolute value of the  parameter $\epsilon_{\mathrm{dd}}$. When $|\epsilon_{\mathrm{dd}}|$ is sufficiently small, the system exhibits a superfluid phase. However, for larger values of $|\epsilon_{\mathrm{dd}}|$, the supersolid phase is favored within a specific range of values of $|\epsilon_{\mathrm{dd}}|$. Beyond this range, the system transitions into the isolated droplet phase.\par
 In the following, we consider a dBEC of $^{164}$Dy atoms with a magnetic moment of $\mu_{m}=9.93\mu_{B}$, where $\mu_{B}$ is the Bohr magneton. The system is examined in the maximally antidipolar regime, with $\phi = \pi/2$. This particular configuration leads to an overall factor of $-1/2$ in Eq.~\eqref{DDI} compared to non-rotating dipoles corresponding to $\phi = 0^{\circ}$. Consequently, the dipolar length becomes $a_{\rm dd}=-65.5a_{0}$, where $a_{0}$ is the Bohr radius. The frequencies of the harmonic potential used in this work are $\omega_x/(2\pi) = 100$~Hz, $\omega_y/(2\pi)= 100$~Hz, and $\omega_z/(2 \pi) = 50$~Hz, resulting in an elongated geometry along the $z$-axis.
 
 \subsection{BdG Approach}\label{bdg_subsec}
In order to unveil the collective excitation spectrum of the system, we perform a Bogoliubov-de Gennes analysis~\cite{pethick, string}. To do so, we perturb the equilibrium solution $\psi_{0}$ using the following ansatz:
\begin{eqnarray}\label{BDG}
& \psi(\mathbf{r},t) =\left\{\psi_0(\mathbf{r}) + \epsilon\left[u(\mathbf{r})e^{-i\Omega t} + v^\ast(\mathbf{r}) e^{i\Omega t}\right] \right\} \nonumber \\ & \times e^{-i\mu t/\hbar}.
\end{eqnarray}
The parameter $\epsilon$ represents a small-amplitude perturbation, while $\mu$ is the chemical potential. The eigenfrequencies and eigenfunctions, represented by $\Omega$ and $(u, v^{*})^{T}$, respectively, are the solutions of the eigenvalue problem resulting from the substitution of Eqs.~\eqref{BDG} into Eqs.~\eqref{eGPE} with terms retained up to the first order in $\epsilon$. Specifically,
 
 \begin{equation}\label{bdg}
 \begin{pmatrix}
 H_\mathrm{s}-\mu + X && X \\
 -X && -(H_\mathrm{s} - \mu + X) 
 \end{pmatrix}
 \begin{pmatrix}
 u \\
 v
 \end{pmatrix}
 = 
 \hbar\Omega
 \begin{pmatrix}
 u \\
 v
 \end{pmatrix},
 \end{equation}
 where $H_s = -\frac{\hbar^2}{2M}\nabla^2 + V(\textbf{r})$ is the single particle Hamiltonian, and the operator $X$ is defined by its action on $q = u,v$ according to
 \begin{equation}
 \begin{aligned}
 Xq(\mathbf{r}) = &\int d \mathbf{r}' U_{\mathrm{dd}}(\mathbf{r}-\mathbf{r}')\psi^{\ast}_0(\mathbf{r}')\psi_0(\mathbf{r})q(\mathbf{r}') \\
 &+g|\psi_0(\mathbf{r})|^2 q(\mathbf{r}) + \frac{3\gamma}{2} |\psi_0(\mathbf{r})|^3 q(\mathbf{r}).
 \end{aligned}
 \end{equation}
 Through the variable transformations $f = (u+v)/\sqrt{2}$ and $g = (u-v)/\sqrt{2}$, Eqs.~\eqref{bdg} can be reduced to two equations with only half the dimensions of the original ones. The resulting reduced equations are then solved by using standard diagonalization methods. 

  \begin{figure}
  	\centering
  	\includegraphics[width = 0.47\textwidth]{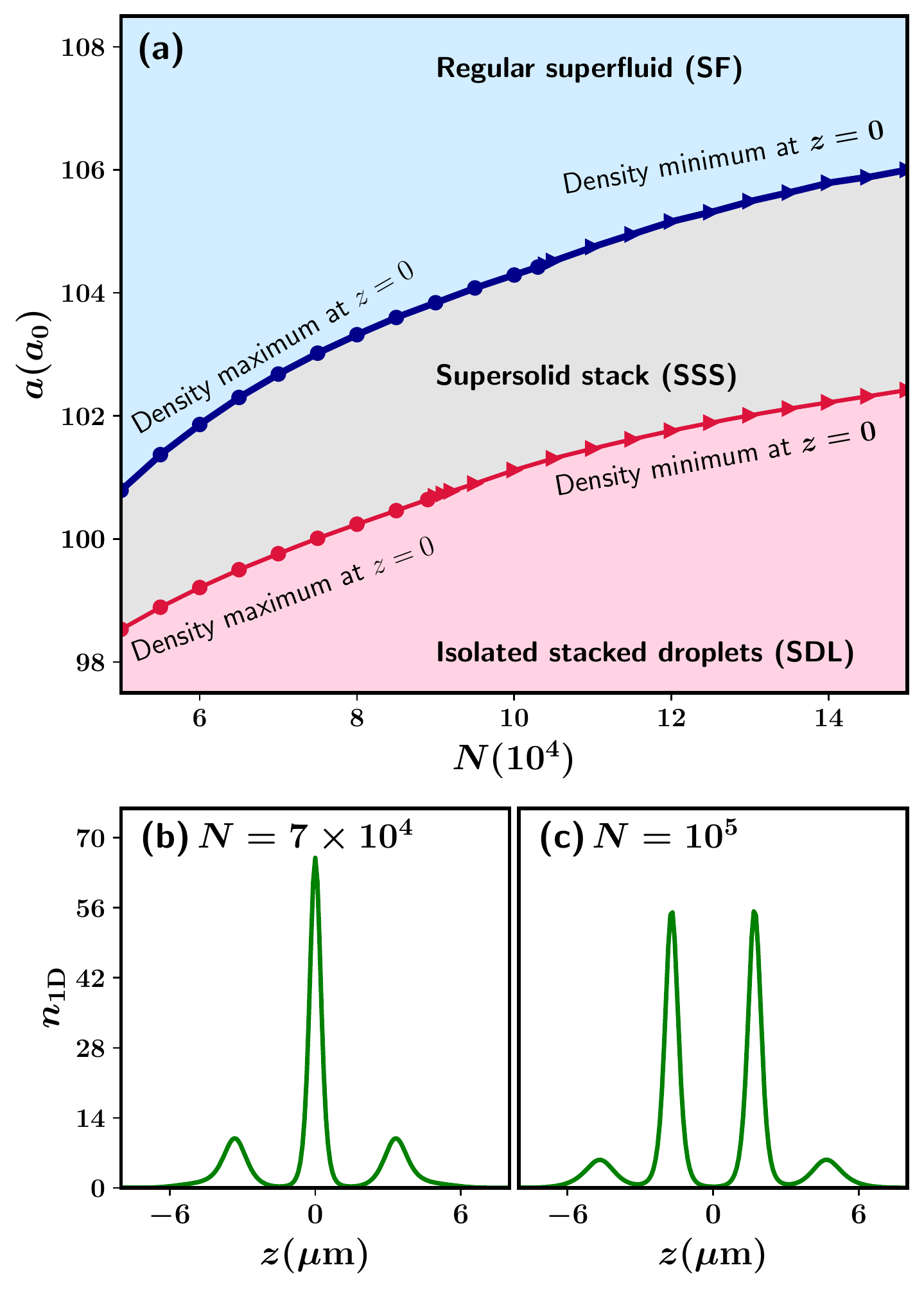}
  	\caption{({\it Color online}) (a) Phases of the antidipolar condensate in terms of the particle number $N$ and scattering length $a$. The thick blue line separates the supersolid disk phases (light grey) from the unmodulated superfluid phase (light blue), which occurs when the contrast $\mathcal{C}=0$. On the other hand, the thin red line divides the supersolid disks from isolated disks (light pink), with reference to contrast $\mathcal{C} > 0.99$. The triangular markers  indicate a phase transition that manifests via a density minimum near the center of the trap, while the circular markers denote the same via a density maximum. The bottom panels show the integrated density profiles $n_{\rm 1D}$ along the $z$-direction for (b) $N=7 \times 10^{4}$ and (c) $N = 10^5$. All other parameters are the same as in Fig.~\ref{den_profile}.}\label{phase_diag}
  \end{figure}
   \begin{figure}
   	\centering
   	\includegraphics[width = 0.48\textwidth]{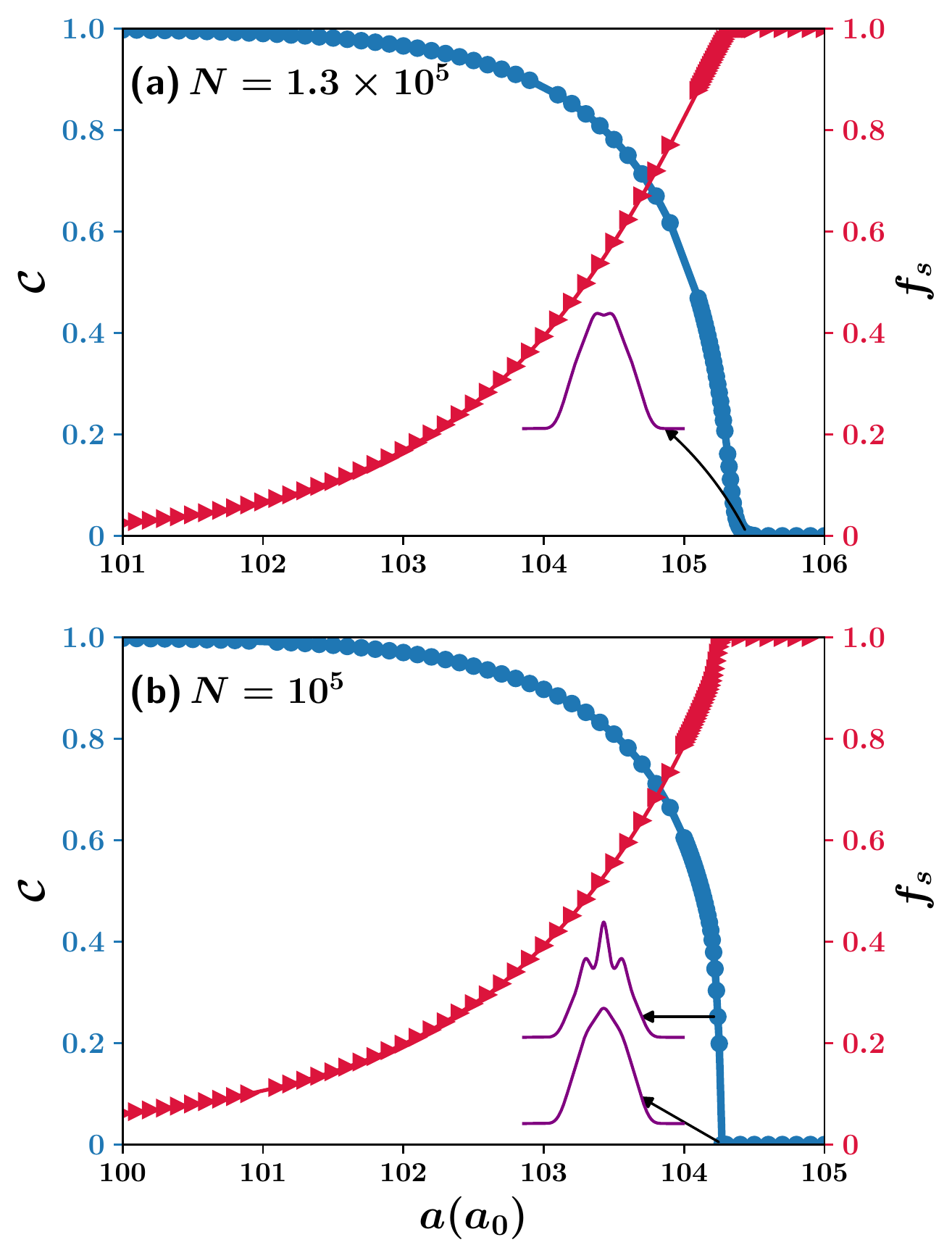}
   	\caption{{\it (Color online)} Ground state contrast $\mathcal{C}$ (blue circular markers) and superfluid fraction $f_s$ (red triangular markers) as function of the scattering length $a$ for two different particle number (a) $N = 1.3 \times 10^{5}$ and (b) $N = 10^{5}$. The insets show the integrated density profiles close to the supersolid and superfluid transitions at scattering length $a=105.45a_{0}$ for $N=1.3 \times 10^{5}$ [(a)], and $a=104.24a_{0}$ and $a=104.27a_{0}$ for $N = 10^{5}$[(b)]. All other parameters are the same as in Fig.~\ref{den_profile}. }\label{contrast_fig}
   \end{figure}
   \begin{figure}[ht]
   	\centering
   	\includegraphics[width = \columnwidth]{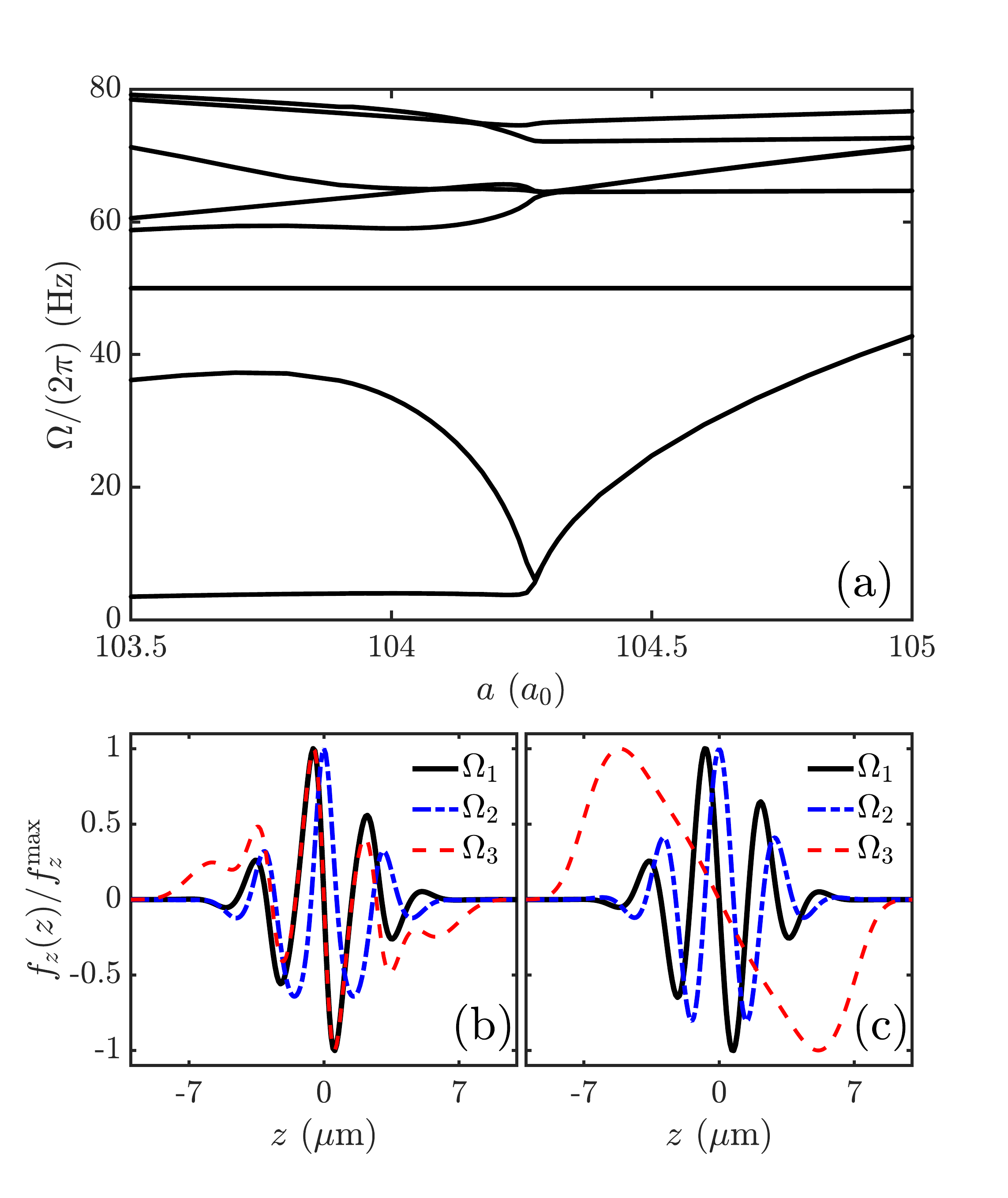}
   	\caption{({\it Color online}) (a) Frequencies $\Omega/(2\pi)$ of the eight lowest excitation modes as a function of the scattering length $a$. The two lower panels show the function $f_z(z) = \int dxdy f(\mathbf{r})$ for the three lowest modes in the (b) supersolid phase for $a = 104a_{0}$, and in the (c) superfluid phase for $a = 104.7a_{0}$, normalized to the maximal value $f_z^\mathrm{max}$. The system consists of $N=10^{5}$ particles, and all other parameters are the same as in Fig.~\ref{den_profile}.}
   	\label{fig_bdg}
   \end{figure}
 \begin{figure*}
 	\centering
 	\includegraphics[width = 0.90\textwidth]{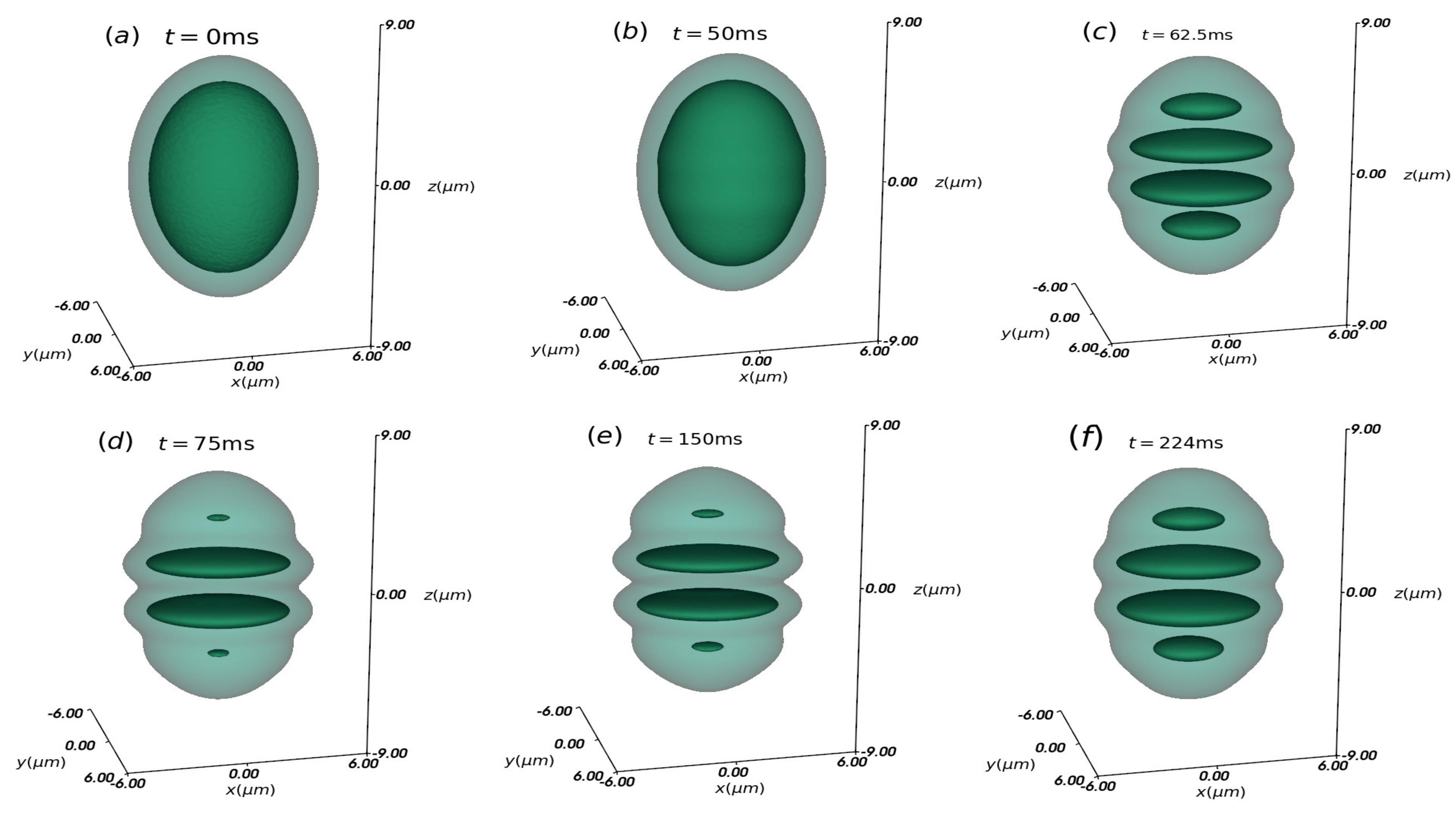}
 	\caption{({\it Color online}) (a)-(b) Dynamical generation of supersolid states consisting of stacks of four disks via an interaction quench. The interaction quench has been performed by linearly reducing the scattering length from $a=110a_{0}$ to $a=105a_{0}$ in $15/\omega_z \approx 47.75 \rm ms$. The density isosurfaces, taken at $20\%$ and $1.5\%$ of the maximum densities, are shown at different instants of time (see the legends). The system is composed of $N = 1.3 \times 10^{5}$ particles with dipolar length $a_{\rm dd}=-65.5a_{0}$ and is confined in a potential with frequencies $(\omega_x, \omega_y, \omega_z)/(2\pi)=(100, 100, 50)\rm Hz$.   }\label{ssIso_dyn}
 \end{figure*}

\section{Ground State properties}\label{GroundState_sec}
Our investigation focuses on exploring the ground-state properties of the antidipolar condensate, with the primary objective of identifying various density structures, determining their region of existence in a phase diagram, and mapping the underlying collective excitation properties during the phase transition. Furthermore, we explore the impact of the number of disks on the sharpness of the phase transition.

\subsection{Identification of Different Phases}\label{phases_subsec}
Figure~\ref{den_profile} displays various ground-state density profiles featuring the unmodulated and modulated phases. For large scattering lengths where the contact interaction dominates the dipolar interaction, a typical superfluid density profile elongated along the $z$-direction is formed, as shown in Fig.~\ref{den_profile}(a) for $N = 1.3 \times 10^{5}$ particles at $a= 110a_{0}$. As the relative strength of the dipolar interaction $|\varepsilon_\mathrm{dd}|$ increases, side-by-side configurations become more favored, leading to their arrangement in the radial plane. However, the presence of a tight radial confinement breaks the condensate into multiple segments along the $z$-direction as the configuration with lowest energy, resulting in the formation of stacked disks. Examples of modulated density structures for two different particle numbers are presented in Figs.~\ref{den_profile}(b)-(c), which demonstrate that the individual disks are connected by a dilute background density, substantiating a SSS phase. Note that the number and distribution of disks along the $z$-direction strongly depend on the total number of particles. \par 

Next, we determine the ground state phase diagram, considering different atom numbers $N$ and scattering lengths $a$ for a fixed $a_{\rm dd} = -65.5a_{0}$. To differentiate between distinct emergent phases, we introduce a contrast measure defined as~\cite{Boettcher2019, Bland2022b}
\begin{eqnarray}
\mathcal{C} = \frac{n_{\rm max} - n_{\rm min}}{n_{\rm max} + n_{\rm min}},
\end{eqnarray}
where $n_{\rm max}$ and $n_{\rm min}$ are the neighboring maximum and minimum densities, respectively. A SF phase lacks periodic density modulation, resulting in $\mathcal{C} = 0$, while a completely isolated density modulated phase corresponds to $\mathcal{C} = 1$. Thus, in the SSS phase, the following relation, $0 < \mathcal{C} < 1$, holds.
The thick blue line in Fig.~\ref{phase_diag} indicates the transition from a regular SF to SSS. Notably, the critical scattering length at which the transition occurs increases with the particle number $N$. Additionally, the number and positions of disks are sensitive to $N$ (and hence density). For instance, an even number of disk-shaped droplets can form along the $z$-direction with a dilute superfluid at $z=0$ (see the triangular markers). This formation occurs for particle number between $1.03 \times 10^{5} < N < 1.04 \times 10^{5}$.  However, for a lower number of particles, one disk is present at $z=0$, leading to an odd number of droplets (see the circular markers). Nevertheless, further decreasing $a$ results in the gradual suppression of the density between individual disks, ultimately rendering them completely isolated from each other. The onset of the SDL phase is indicated in Fig.~\ref{phase_diag} by the thin red line. This transition also manifests by a density maximum (circular markers) or minimum (triangular markers) at $z=0$, depending on particle number. In our current setup, we notice that an odd number of disks with one situated at the center form for  particle numbers between $9 \times 10^4 < N < 9.1 \times 10^{4}$. Typical examples for isolated three- and four-disk stacks  are demonstrated for $N = 7 \times 10^{4}$ and $N = 10^{5}$ particles, respectively, in Figs.~\ref{phase_diag}(b)-(c), showing integrated density profiles defined as $n_{\rm 1D}(z) = \int dx dy \abs{\psi(x, y, z)}^2$. This observable, which we also utilize later,  is experimentally detectable, e.g. via in-situ imaging~\cite{ Hertkorn2021b, Chomaz2022}. \par 

To gain a deeper understanding of the nature of the transition, we investigate the contrast $\mathcal{C}$ as a function of the scattering length $a$ for two different particle numbers; see Fig.~\ref{contrast_fig}. In addition, we use Leggett's upper bound to estimate the superfluid fraction $f_s$~\cite{Leggett1970, Leggett1998} in the central region of length $2L$\footnote{The length $2L$ spans the central region between the first two side-minima for three disks and the two first-side maxima for four disks, along the $z$-axis~\cite{Scheiermann2023}.}.
\begin{eqnarray}\label{legget_SF}
f_s =\frac{(2L)^2}{N}\left[ \int dz \frac{1}{\int dx dy \abs{\psi}^2}\right]^{-1}.
\end{eqnarray}
For $N = 1.3 \times 10^{5}$ particles, we observe the continuous formation of  disks on both sides of $z=0$ as $\mathcal{C}(f_s)$ gradually approaches zero (one), indicating the formation of modulated density structures with a strong superfluid connection between them. As the scattering length $a$ decreases, we observe that $\mathcal{C}$($f_s$) increases (decreases) due to a pronounced density modulation. At $a=102.1a_{0}$, the disks become completely isolated from each other and $\mathcal{C}>0.99$. Interestingly, for $N = 10^{5}$ particles, the superfluid to supersolid transition is much sharper, but it still remains continuous. We observe a disk formation at $z=0$ for this transition, unlike in the case of $N=1.3 \times 10^{5}$ particles. The contrast drops from $\mathcal{C}=0.252$ at $a=104.24a_{0}$ to $\mathcal{C}=0$ at $a=104.27a_{0}$. Our results are consistent with the findings of Ref.~\cite{Zhang2019}, where the supersolid phase transition is argued to be continuous at larger particle numbers closer to the thermodynamic limit due to insignificant kinetic energy contributions.   A close inspection also reveals that the superfluid state just before the transition to the supersolid state does not exhibit the typical parabolic density. Nonetheless, they are still classified as superfluid since they demonstrate a non-periodic density modulation. For instance, for $N=1.3 \times 10^{5}$, $n_{\rm 1D}$ has a tendency to develop flat top [Fig.~\ref{contrast_fig}] profile, while for $N=10^{5}$, a sharp peak appears [Fig.~\ref{contrast_fig}(b)]. Turning to the SDL side of the curves, we observe that the contrast $C$ gradually approaches zero as the system enters the isolated droplet regime for both $N = 1.3 \times 10^{5}$ and $N = 10^{5}$ particles. Notably, our numerical simulations show that the superfluid fraction, as calculated by Eq.~\eqref{legget_SF}, does not completely vanish during the SSS-SDL transition. Additionally, $f_s$ gradually decreases and becomes $f_s < 0.1$ when $\mathcal{C} > 0.99$, indicating the transition to the SDL phase. This observation is consistent with the findings of Ref.~\cite{Kirkby2023}.

\subsection{Collective Excitation Spectra}\label{excitation_subsec}
Having identified the ground-state phase diagram and showcased some density profiles, we turn to  study the behavior of the collective excitations. The low-lying collective excitation spectra for the supersolid phase in regular dipolar condensate have been reported in Refs.~\cite{Chomaz2018,Natale2019, Hertkorn2019,Schmidt2021, Hertkorn2021b, Hertkorn2021c, Buehler2022}. To accomplish this for the antidipolar condensate, we fix the particle number at $N=10^5$ and compute the spectrum over a range of values of the $s$-wave scattering length by solving the BdG Eqs.~\eqref{BDG}. The obtained results are displayed in Fig.~\ref{fig_bdg}. Specifically, we have examined the low-lying excitations that are primarily related to excitations in the $z$-direction. The frequencies of the eight lowest modes are presented in Fig.~\ref{fig_bdg}(a). We notice that as the scattering length is decreased from $a=105a_0$, the two lowest (degenerate) modes eventually split into two around $a\sim104.26a_0$, marking the transition from the superfluid to the supersolid phase. It is interesting to note that the point at which the degeneracy breaks coincides with the point where the contrast drops to zero, as shown in Fig.\ref{contrast_fig}(b). This implies that the breaking of degeneracy can be used as a precise definition of the transition point from the superfluid to the supersolid phase. This behavior is analogous to what has been observed for a regular dipolar supersolid in an elongated harmonic trap~\cite{Hertkorn2019}.
To investigate the nature of these modes further, we note that to the lowest non-vanishing order in $\varepsilon$, the density can be written as
\begin{equation}
n(\mathbf{r}, t) = \psi_0^2(\mathbf{r}) + 2\sqrt{2} \varepsilon f(\mathbf{r})\psi_0(\mathbf{r})\cos(\Omega t).
\end{equation}
The time-evolution of the density associated with a mode is consequently characterized by the function $f$, and we display the integrated version $f_z(z) = \int dxdy f(\mathbf{r})$ of the three lowest modes in Figs.~\ref{fig_bdg}(b)-(c). The modes in the SSS phase at $a = 104a_{0}$ are shown in Fig.~\ref{fig_bdg}(b), while the modes in the superfluid phase at $a=104.7a_{0}$ are shown in Fig.~\ref{fig_bdg}(c). Interestingly, the two lowest modes are very similar in both phases, although their effects are quite different due to their different spatial distributions. For the supersolid, the lowest mode corresponds to a center of mass excitation where the high-density regions move back and forth. Thus it is a Nambu-Goldstone mode associated with the spontaneous breaking of translation symmetry~\cite{Nambu2009}. The second-lowest mode is a Higgs (amplitude) mode~\cite{Pekker2015}, where the crystalline regions periodically increase and decrease in density in a fashion opposite to that of the superfluid background. The third-lowest mode is the dipole mode, which has a constant frequency $\Omega/(2\pi) = 50$ Hz, independent of the interaction strength~\cite{Kohn1961}. The dipole mode remains decoupled from other modes, thereby serving as a gauge for numerical accuracy. In the superfluid phase [Figs.~\ref{fig_bdg}(c)], the first two modes, $\Omega_1$ and $\Omega_2$, have the same frequency, and the third mode is the dipole mode. This dipole mode changes character significantly as the system transitions from a superfluid to a supersolid. The function $f_z$ of the dipole mode in the supersolid phase has a very prominent dipole-like structure that is composed of a single positive and negative region, see the red dashed curve in Fig.~\ref{fig_bdg}(c). This mode in the supersolid phase [red dashed line, $\Omega_3$, in the Fig.~\ref{fig_bdg}(b)] acquires a behaviour that appears to be a mix of its lowest mode [solid black line, $\Omega_1$, in Fig.~\ref{fig_bdg}(b)] and the dipole mode in the superfluid phase [$\Omega_3$ in Fig.~\ref{fig_bdg}(c)], possessing multiple local minima and maxima. This mode is often attributed to in-phase (between the crystal and the superfluid background) centre-of-mass oscillation, referred to as the in-phase Goldstone mode~\cite{Guo2019}. 
 
\section{Quench Dynamics across the phase transition}\label{Dyn_sec}

After explicating the supersolid state and its underlying collective excitation spectrum within the ground-state phase diagram, we now shift our focus to studying the non-equilibrium dynamics of our system. To trigger these dynamics, we will vary the scattering length from an initial value of $a=a_{i}$ to a final value of $a=a_f$.   A particular focus will be on examining how the layers of disks can be generated dynamically. We will also illustrate the emergence of excitations that manifests as density oscillations and how the excitation frequencies relate to those calculated via  the BdG analysis at $a=a_f$. 

\subsection{Dynamical Generation of Supersolid Disks}\label{dyn_density_profile}

\begin{figure}
	\centering
	\includegraphics[width = 0.46\textwidth]{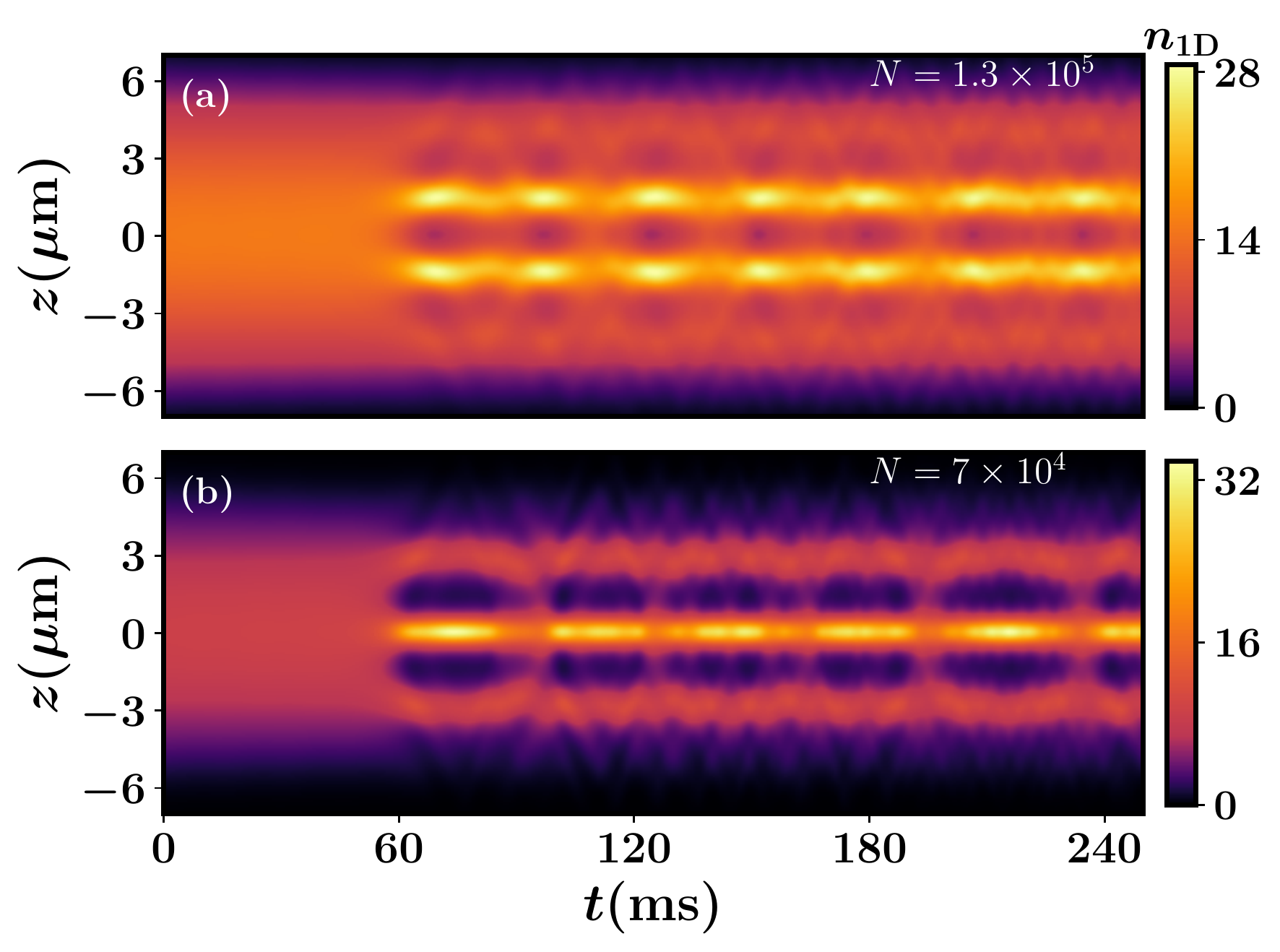}
	\caption{(Color online) Generation of supersolid states for two different particle numbers (a) $N= 1.3 \times 10^{5}$, and (b) $N = 7 \times 10^{4}$, utilizing the dynamic ramp of the scattering length $a$ from an initial value $a=110a_{0}$ to the final value $a=105a_{0}$[(a)], and $a=102a_{0}$[(b)]. The colorbar represents the integrated density $n_{\rm 1D}$ in unit of $1000$~$\mu \mathrm{m}^{-1}$. The duration of dynamic ramp is $15/\omega_z \approx 47.75$~ms. All other parameters are the same as in Fig.~\ref{den_profile}.  }\label{supsol_dyn}
\end{figure}

We present our results for a system comprising $N=1.3 \times 10^{5}$ particles by showing the 3D density isosurfaces in Figs.~\ref{ssIso_dyn}(a)-(f). Our investigation begins with an initial state prepared at $a_{i}=110a_{0}$. We then introduce a very small amplitude noise [Figs.~\ref{ssIso_dyn}(a)] to the ground state, followed by a gradual decrease of the scattering length to $a_f = 105a_{0}$ over a time span of $47.75$~ms\footnote{This time span is chosen such that we can adiabatically produce the states which have the same number of droplets as those in the ground state of the system. A sudden quench would typically produce more droplets than the ground state~\cite{Boettcher2019}.}. As the system dynamically enters the SSS phase within the ground-state phase diagram [Fig.~\ref{phase_diag}(a)], a modulational instability sets in, resulting in the onset of density modulation. This can be observed by careful inspection in Fig.~\ref{ssIso_dyn}(b). Subsequently, for $t > 47.75$~ms, the system quickly forms four prominent circular disks arranged in layers, connected by a dilute superfluid background as depicted in Fig.~\ref{ssIso_dyn}(c). A more precise analysis of the temporally resolved dynamics leading to the SSS phase can be done by invoking the integrated density $n_{\rm 1D}(z,t)$. The time evolution of $n_{\rm 1D}$ for two different particle numbers is shown in Fig.~\ref{supsol_dyn}. Indeed, the formation of four disks, with the central two having the highest densities from an initial non-modulated state, is evident in Fig.~\ref{supsol_dyn}(a), corresponding to $N = 1.3 \times 10^{5}$ particles. Similarly, an odd number of disks can be formed dynamically. An example of the formation of a state with three disks is illustrated in Fig.~\ref{supsol_dyn}(b) for $N =7 \times 10^{4}$. The final scattering length $a_f = 102a_{0}$ is achieved from an initial $a_{i}=110a_{0}$ in the same time span of $47.75$~ms as before. The superfluid connection remains more robust for $N = 1.3 \times 10^{5}$ particles when compared to $N = 7 \times 10^{4}$. We remark that the value of contrast at the ground state corresponds to $\mathcal{C} = 0.356 $ at $a =105a_{0}$ for $N = 1.3 \times 10^{5}$, and $\mathcal{C} = 0.54$ at $a =102a_{0}$ for $N = 7 \times 10^{4}$, implying a weaker background for the latter, which also becomes evident in dynamically generated SSS phase [see Figs.~\ref{supsol_dyn}(a)-(b)].
Another important observation from  Figs.~\ref{ssIso_dyn}(d)-(f) is that the thickness of each individual crystal changes during the dynamics, implying particle flow among them as well as the emergence of density oscillations. This particle flow causes a persistent alteration of the density, radial width, and thickness [see Figs.~\ref{supsol_dyn}(a)-(f)], thereby triggering compressional dynamics, which we will expound on next.

\begin{figure}
	\centering
	\includegraphics[width = 0.46\textwidth]{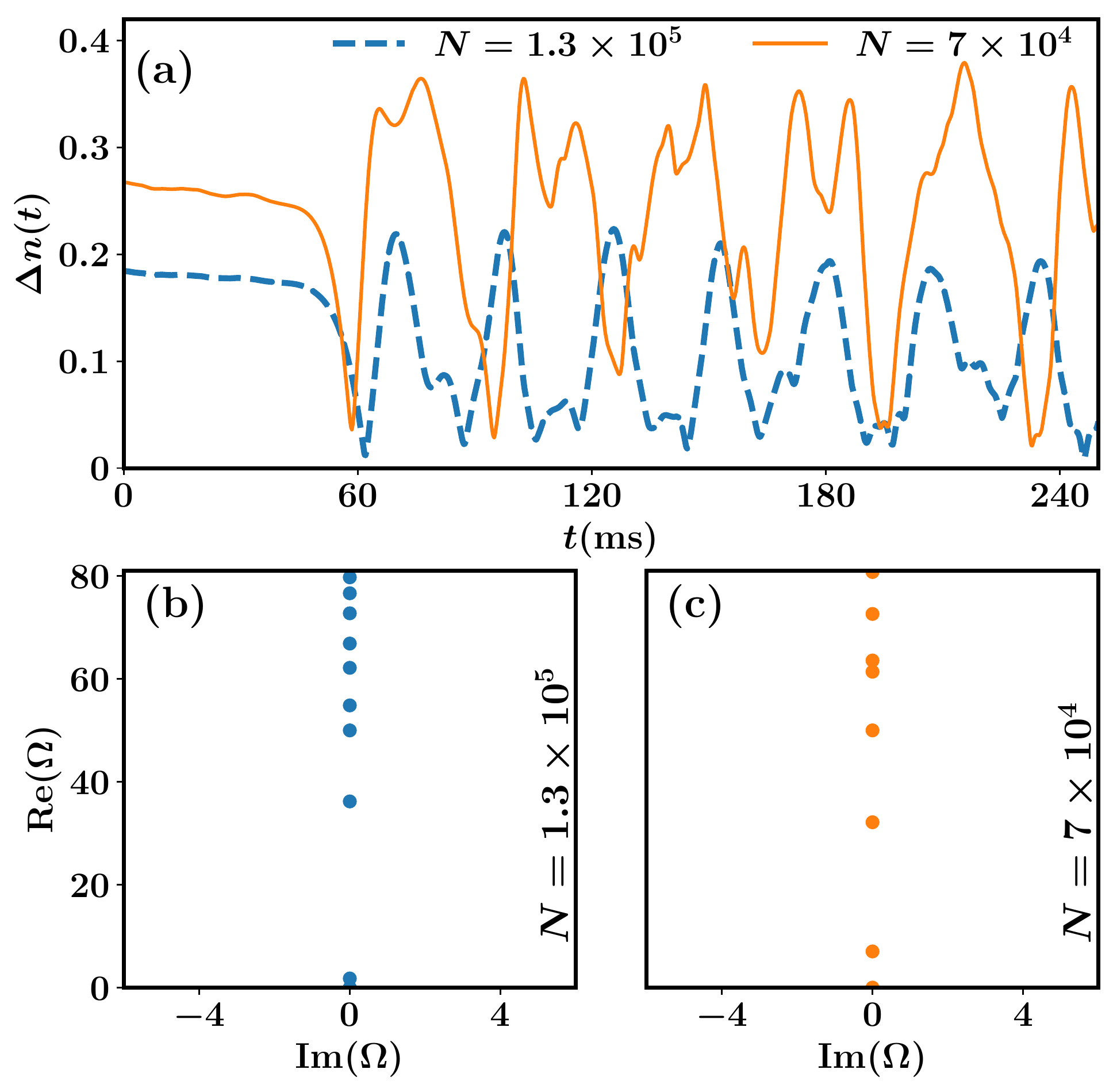}
	\caption{(Color online)	(a) Time evolution of the integrated density difference with respect to the ground state $\psi_{0}$ at the final scattering lengths is quantified via $\Delta n(t) = (1/N)\int d\vb{r}(\abs{\psi(\vb{r};t)}^2 - \abs{\psi_{0}{(\vb{r})}}^2)$ for two different particle numbers (see the legends). The dynamics is triggered by ramping the scattering length from $a=110a_{0}$ to $a=105a_{0}$ ($N = 1.3 \times 10^{5}$), and $a=102a_{0} (N = 7 \times 10^{4})$. The real and imaginary parts of the collective excitation spectra $\Omega$ are presented in (b) for $N = 1.3 \times 10^{5}$ and (c) for $N = 7 \times 10^{4}$, calculated by BdG analysis at $a=105a_{0}$ and $a=102a_{0}$, respectively. The BdG frequencies are shown in units of $2\pi\times$Hz.}\label{den_osc}
\end{figure}
\subsection{Collective Oscillation During Dynamics}\label{collect_osc_dyn}

Here we  evaluate the extent to which the density of the system during the time evolution remains similar to that of the ground state at the same scattering length. This can be monitored by analyzing the observable $\Delta n(t)$, as given by
\begin{eqnarray}
\Delta n(t) = (1/N)\int d\vb{r}(\abs{\psi(\vb{r},t)}^2 - \abs{\psi_{0}{(\vb{r})}}^2),
\end{eqnarray}
where $\psi_{0}$ represents the ground state wavefunction at $a_f$. When the value of $\Delta n(t)$ approaches zero, it indicates that the dynamically generated supersolid state is very similar to the ground state. However, any deviation from zero arises due to the collective excitations triggered during the dynamics. The behavior of $\Delta n(t)$, which relates to the density evolution displayed in Fig.~\ref{supsol_dyn}(a)-(b), has been presented in Fig.~\ref{den_osc}(a). Note that $\Delta n(t = 0)$ is higher for $N =1.5 \times 10^{5}$ than for $N =7 \times 10^{4}$ due to a stronger superfluid connection in the final state, $\psi_{0}$, of the former, making it much closer to the initial superfluid ground state.
As the scattering length is gradually ramped down to the final value, $\Delta n(t)$ gradually decreases and eventually reaches a minimum, as shown in Fig.~\ref{den_osc}(a). This minimum value indicates the generation of a density-modulated state that closely resembles the ground state at the final scattering length $a_f$. Subsequently, $\Delta n(t)$ exhibits an oscillatory behaviour that involves multiple oscillation frequencies.
The oscillation amplitude is larger for $N = 7 \times 10^{4}$ than $N = 1.3 \times 10^{5}$ because the change in scattering length $a_{f} - a_{i}$ is larger in the same time span, making the quench less adiabatic in the former case. As mentioned earlier, such oscillations in $\Delta n(t)$ stem from the continuous particle flow between the crystal and background, leading to compressional dynamics. The two dominant frequencies of oscillations are calculated by performing a Fourier transformation of $\Delta n(t)$\footnote{To compute the frequencies, we have propagated $\Delta n(t)$ until the time $T_f=1$~s, and as a result, the maximum frequency error is estimated to be $1/T_f = 1$~Hz.}, yielding $\omega_{f_1}/(2\pi)\approx \rm 36 \pm 0.5 Hz$ and $\omega_{f_2}/(2\pi)\approx \rm 72 \pm 0.5 Hz$ for $N =1.3 \times 10^{5}$, and $\omega_{f_1}/(2\pi)\approx \rm 28 \pm 0.5 Hz$ and $\omega_{f_2}/(2\pi)\approx \rm 72 \pm 0.5 Hz$ for $N =7 \times 10^{4}$. Next, we check whether these frequencies are related to the low-lying excitations of the ground state at the final scattering length, $a_{f}$. It should be noted that the excitation spectra at scattering lengths other than the final value may also be triggered during the quenching process. However, their contribution to the oscillation of $\Delta n(t)$ is expected to be negligible, as the system spends most of the time at $a_f$. The frequencies of the low-lying collective excitation spectra calculated at $a=105a_0$ and $a=102a_0$ are presented in Figs.~\ref{den_osc}(b) and (c) for $N = 1.3 \times 10^{5}$ and $N =7 \times 10^{4}$, respectively. Notably, the dominant oscillation frequency in $\Delta n(t)$ is very close to the second frequency (excluding the zero frequency) that corresponds to the Higgs amplitude mode, as depicted in Figs.~\ref{den_osc}(b)-(c). Additionally, the second dominant frequency close to $72 \rm Hz$ is also evident in the spectrum. Finally, it should be noted that we have illustrated the intricate connection discussed above by considering these two particle numbers as representative of three- and four-crystal states. However, the discussion is generally valid for all states in the SSS phase, where the non-equilibrium dynamics induced by the quench indeed emanates from triggering the low-lying excitations of the system.

\begin{figure}
	\centering
	\includegraphics[width = 0.46\textwidth]{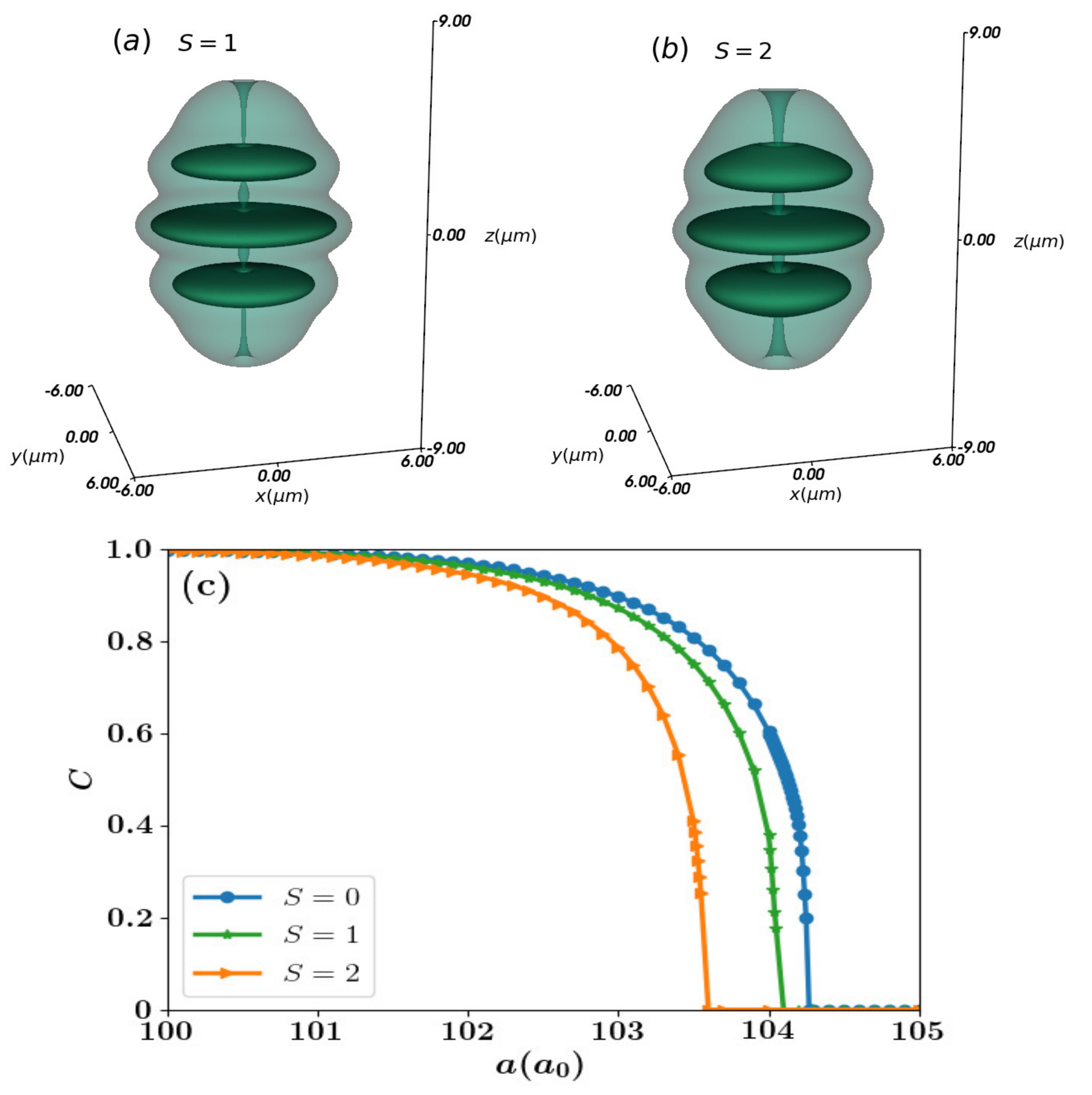}
	\caption{(Color online) Density isosurfaces within a supersolid state in the presence of a vortex line with charge (a) $S=1$ and (b) $S=2$. (c) The contrast $\mathcal{C}$ as a function of scattering length $a$ in the presence ($S=1, S=2$) or absence ($S=0$) of a vortex line. The density isosurfaces represent $20\%$ and $2.5\%$ of the maximum density. The $^{164}$Dy system  is composed of $N = 10^{5}$ particles with dipolar length $a_{\rm dd}=-65.5a_{0}$, and confined in a traping potential with frequencies $(\omega_x, \omega_y, \omega_z)/(2\pi)=(100, 100, 50)$~Hz. }\label{SS_vortex}
\end{figure}

\section{Impact of a vortex line}\label{vortex_sec}

Let us finally briefly illustrate the effect of a non-linear defect on the formation of multiple-disk structures and the associated superfluid background. As the disks are relatively flat structures, they offer an opportunity to investigate quasi-2D physics, particularly in relation to vortices. Hence, it is crucial to examine the influence of vortex line penetration on the phase transition. To realize a vortex line of charge $S$ passing through the center of the disks, we perform the following transformation
\begin{eqnarray}
\psi(x, y, z) = \psi(x, y, z)e^{i S \tan^{-1}(y/x)}
\end{eqnarray}
to the wavefunction during the imaginary time evolution, which ensures a vanishing density at the center.\par 
Figures \ref{SS_vortex}(a) and \ref{SS_vortex}(b) depict the 3D isosurfaces representing the supersolid state for $S=1$ and $S=2$, respectively, realized at $a=103a_{0}$ for $N=10^5$ particles. The vortex line has a larger radial thickness in the dilute superfluid background than in the localized crystal structure, which varies between $S=1$ and $S=2$. This suggests that the superfluid connection between the crystal changes due to the presence of a vortex. To investigate this further, we explore whether imprinting a vortex line changes the supersolid region, as shown in Fig.~\ref{SS_vortex}. The contrast $\mathcal{C}$ is plotted as a function of the scattering length $a$ in both the presence and absence of a vortex line. The results show that a higher-charge vortex shifts the supersolid phase towards a lower scattering length compared to the vortex-free system. However, the contrast asymptotically approaches each other towards the isolated droplet phases, indicating that the SSS to SDL transition is not affected by the presence of a vortex. Between the SF and SSS phase, the contrast decreases sharply towards lower values for higher vortex charges. It is energetically costly for a vortex to dig a hole within the highly localized crystal compared to the dilute superfluid, and therefore it increases superfluid connection, minimizing the crystal density. \par  

In addition, we remark that we have analyzed the stability of vortex lines in real-time dynamics. It is observed that a single charge vortex remains stable, but the double-unit charged vortex line eventually bends and breaks. However, the detailed discussion of their dynamics is beyond the scope of this manuscript. 

 \section{Conclusions}\label{conclusions} 
In this paper, we have reported a novel supersolid state formed in the form of stacked disk-like droplets connected by a dilute superfluid  in an antidipolar condensate~\cite{WenzelPhD}. Considering an elongated geometry with experimentally relevant trapping frequencies we have presented a phase diagram in the  parameter space of particle number $N$ and scattering length $a$, identifying the regions of existence for a superfluid, supersolid disks, and isolated disks. \par 
We have utilized a contrast measure and superfluid fraction to differentiate between different emergent phases. As the scattering length decreases, the contrast (superfluid fraction) increases (decreases) due to a pronounced density modulation, leading to the formation of disk-shaped density modulations. The critical scattering length for the transition from a regular superfluid to a  supersolid stack phase increases with the particle number. The number and distribution of disks along the $z$-direction are strongly dependent on the total number of particles.\par 
Building upon the phase diagram, we have studied the collective excitation spectrum across the superfluid-supersolid phase transition in the antidipolar condensate. Specifically, we have computed the frequency of the eight lowest modes and characterized the density of the first three modes. The breaking of degeneracy of the two lowest-lying modes marks the transition point from the superfluid to the  supersolid stack phase. Additionally, the third lowest mode, the so-called dipole mode, also changes significantly as the system undergoes a phase transition. We have also examined the influence of vortex line penetration on the phase transition, with the supersolid region shifting towards weaker contact interaction upon imprinting a vortex line.\par
As a next step, we have studied the non-equilibrium dynamics triggered by ramping down the scattering length across the superfluid and supersolid phase transition. We have demonstrated the dynamic generation of four and three disks connected by a dilute superfluid, considering two different particle numbers. The particle flow between the disks triggers a collective oscillation in the density. We have identified two dominant frequencies of oscillations, which can also be found in the low-lying collective excitation spectra of the ground state at the final scattering length.\par
The present work has opened up numerous promising research directions for future endeavors. To further enhance our understanding, it would be fascinating to investigate the response of the disks and background superfluid to external rotation and dynamically observe how vortex lines penetrate through the disk-like density modulations. Currently, vortices are being sought actively in dipolar supersolids, and the recent creation of vortices in an unmodulated dipolar condensate presents optimistic prospects in this area~\cite{Klaus2022, Bland2023}.
The 2D nature of the crystal makes the antidipolar condensate an intriguing setup along that direction. Further exploration will be undertaken to examine  supersolid stacks and isolated stacked droplets in the context of binary mixtures~\cite{Arazo2023}. Direct formation of such a supersolid structure via evaporative cooling along the line of Ref.~\cite{Bland2022} and investigating the thermal properties~\cite{Baena2023} would be highly interesting.
\section*{Acknowledgements} This work was financially supported by the Knut and Alice Wallenberg Foundation and the Swedish Research Council. Many valuable inputs from  P. St\"{u}rmer and L. Chergui are gratefully acknowledged.

\appendix

\twocolumngrid

\section{Three-Body Atom Loss}\label{threeBody_appen}
\begin{figure}
	\centering
	\includegraphics[width = 0.46\textwidth]{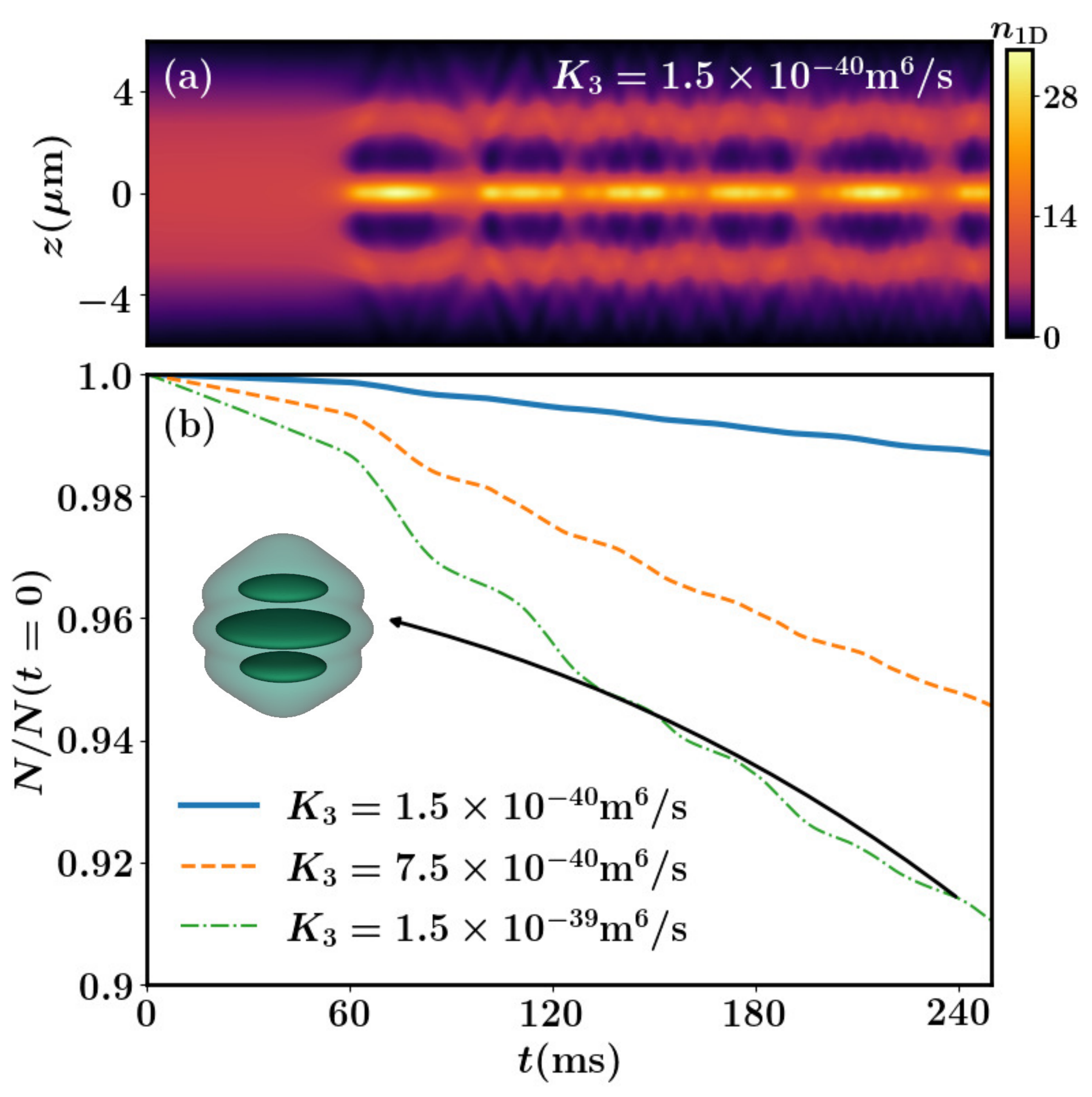}
	\caption{(Color online) (a) The time evolution of the integrated density $n_{\rm 1D}$ for a quench from an unmodulated state to a supersolid state, including three-body atom loss with a three-body coefficient $K_3 = 1.5 \times 10^{-40}\text{~}\rm m^6/s$. The quench has been performed by ramping off the scattering length from $a=110a_{0}$ to $a=102a_{0}$ in $48$~ms. (b) The time evolution of the atom number $N$ for different loss coefficients (see the legends), considering the same quench protocol. The system consists of $N=70000$ $^{164}$Dy atoms confined in a harmonic trapping potential with frequencies $(\omega_x, \omega_y, \omega_z)/(2\pi)=(100, 100, 50)\text{~}\rm Hz$, with $a_{\rm dd}=-65.5a_0$. The colorbar represents the density in units of $1000 \text{~}\mu\mathrm{m}^{-1}$	}\label{loss_fig}
\end{figure}

In experiments, the lifetime of the supersolid state is limited by three-body losses, which has been previously discussed in the context of dBECs with positive dipolar lengths~\cite{Tanzi2019a,Chomaz2019,Boettcher2019}. This type of losses is caused by highly localized density arrays, making it crucial to understand how it affects the formation and persistence of the SSS in anti-dBECs. A comprehensive understanding of this phenomenon is necessary for the successful experimental realization of SSS states. In simulations, the loss can be modeled by adding an imaginary contribution, represented by $(-i\hbar K_3/2)\abs{\psi(\textbf{r})}^4$, to Eq.~\eqref{eGPE}, where $K_3$ denotes the three-body recombination rate.

To demonstrate our findings, we examine the formation of a supersolid state in a system composed of $N=70000$ $^{164}$Dy atoms by reducing the scattering length from $a=110a_{0}$ to $a=102a_{0}$ over a period of $47.78$~ms. Realistic three-body losses are considered in simulations using the experimentally measured loss coefficient, $K_3 = 1.5 \times 10^{-40}\ \rm m^{6}/s$~\cite{Boettcher2019}. Fig.~\ref{loss_fig}(a) shows the time evolution of the integrated density $n_{\rm 1D}(z)$. The maximum atom loss is $\approx 5.6\%$ of the initial population in $250$~ms, as displayed by the blue line in Fig.~\ref{loss_fig}(b). After $t>47.75$~ms, the initially unmodulated state deforms into three disks. Despite continuous population loss, see Fig.~\ref{loss_fig}(b), a dilute background density always connects the disks throughout the dynamics. We have also investigated the population loss and supersolid state generation with larger three-body loss coefficients, $K_3 = 7.5 \times 10^{-40} \text{~}\rm m^{6}/s$ and $K_3 = 1.5 \times 10^{-39} \text{~}\rm m^{6}/s$, respectively. As expected, larger $K_3$ leads to more atom losses, and $N(t)$ exhibits oscillatory behavior [Fig.~\ref{loss_fig}(b)]. This tendency is caused by the atom losses being influenced by the severe changes in density distribution (via the term $\abs{\psi(\vb{r})}^4$), which also undergoes periodic expansion and contraction after the quench. Nonetheless, the supersolid state remains robust in long-time dynamics even for $K_3 = 1.5 \times 10^{-39}\text{~}\rm m^{6}/s$, as evidenced, for example, by the density isosurface at time $t=240\text{~}\rm ms$.

\section{Computational Details}\label{comdetails}

Here, we provide a detailed account of the numerical simulations used to obtain the results described in the main text.
The extended Gross-Pitaevskii equation, Eq.~\eqref{eGPE}, is cast into a dimensionless form in our simulations by rescaling the length, the time in terms of the harmonic oscillator length scale $l_{\rm osc} = \sqrt{\hbar/ m\omega_z}=1.11 \text{~}\rm \mu m$, and the trap frequency $\omega_z$, respectively. Then the wavefunction is scaled accordingly as $\psi(\vb{r}',t) = \sqrt{l^3_{\rm osc}/N} \psi(\vb{r},t)$. We solve the resulting dimensionless equation using the split-step Crank-Nicholson method, see Ref.\cite{crank_nicolson1947}. The stationary (lowest energy) states of the dBEC are obtained through imaginary time propagation. Furthermore, we apply the transformation $\frac{\psi(\vb{r}, t)}{\norm{\psi(\vb{r}, t)}} \rightarrow 1$ at each imaginary time-step  of this procedure. This preserves the normalization of the wavefunction, while convergence is reached until relative deviations of the wave function (at every grid point) and energy between consecutive time-steps are smaller than $10^{-5}$ and $10^{-7}$, respectively. This solution is then used as an initial state for the  dynamical simulations, where the eGPE is propagated in real time. It should be noted that calculating the stationary state solution of Eq.~\eqref{eGPE} is an involved task due to many close-lying local minima in the energy surface, which necessitates extensive sampling over many different initial conditions to identify the most probable lowest-energy solutions.
Our simulations are carried out in a 3D box characterized by a grid $n_x \times n_y \times n_z$ corresponding to $(128 \times 128 \times 256)$. The employed spatial discretization (grid spacing) refers to $\Delta_x =0.12l_{\rm osc} $, $\Delta_y=0.12l_{\rm osc}$, and $\Delta_z =0.08l_{\rm osc}$, while the time step of the numerical integration is $\Delta_t = 10^{-5}/\omega_z$.

\bibliographystyle{apsrev4-1.bst}
\bibliography{reference.bib}

\end{document}